# HOW CAN WE DISTINGUISH TRANSIENT PULSARS FROM SETI BEACONS?

James Benford and Dominic Benford
Microwave Sciences
Lafayette, CA

How would observers differentiate SETI beacons from pulsars or other exotic sources, in light of likely beacon observables? Bandwidth, pulse width and frequency may be distinguishing features. We show a method of analyzing an observed radio transient in terms of a possible beacon. Such transients are seen pulsar searches, and could be evidence of civilizations higher than us on the Kardashev scale. Galactic-scale beacons require resources larger than earth presently has available. From the examples, a civilization lower on the Kardashev scale will have a narrower beam, revisit less frequently, and so will be harder to observe. But lower power beacons will probably be more numerous, so we should learn how to identify them. We urge observers to consider SETI beacons as a candidate explanation when perplexing non-repeating signals are seen in the radio sky.

Pulsars are clearly radiation from rotating neutron star magnetospheres. There is increasing interest in certain transient phenomena of uncertain origin, but likely due to atypical pulsars[1]. These are occasional observations of repetitive sources, some re-observed, some not.

Such observations have another possible origin: extraterrestrial Beacons trying to attract attention from civilizations such as ourselves. We have recently described such Beacons in terms of how *our* civilization would build large broadcasting transmitters if we were to undertake announcement of our existence (METI, Messaging to Extraterrestrial Intelligences) or even detailed communication[2]. Our approach uses cost optimized beacons: designing under a likely constraint on cost affects design choices and therefore observable parameters. We found that beacons are likely to be pulsed, both to lower the cost and to make the signal more noticeable. In fact, beacon might mimic a pulsar, because they are likely to be studied. Of course, they would try to distinguish themselves in some way, such as modulations of amplitude, frequency hopping, etc. They would not be isotropic, as pulsars are not isotropic, in order to lower cost. So they might be much like a lighthouse, sweeping over a region of the sky, scanning in a raster pattern, close to the galactic plane. The number of pulses an observer would see must be 'enough' to distinguish it: large enough to contain the modulation, but small enough to allow scanning of the beam over the designated target region ('dwell' time), after which it would return ('revisit' time).

There is increasing interest in shorter transient astronomical sources, so researchers should be aware of the likely properties of beacons. How can observers distinguish such beacons from pulsars or other exotic sources[3]? Here I consider transient observations in light of likely beacon observables, with examples of how to analyze observational data in terms of possibly being a beacon, deducing beacon parameters.

**Factors to distinguish Bacons from pulsars are:**

**Bandwidth** A quick check for pulsar behavior is to consider the power law spectrum. An obvious difference is that pulsars have large bandwidths, typically peaking at ~1 GHz, falling with about the square of frequency. So a half-power bandwidth would be ~400 MHz. This is much larger than our powerful microwave devices, which have much smaller bandwidths. That is because they are based on conversion of electron-beam energy to microwaves by techniques using resonance. For such devices, there is a tradeoff, called the gain-bandwidth product, which means that higher efficiency in microwave generation comes when there is high gain and small bandwidth. (Here 'small' means much larger than the 1 Hz signals many SETI listeners have searched for, though.) Efficient Earth sources have bandwidths from 100 kHz to few MHz in the ~1GHz range where pulsars radiate most of their energy, and efficiencies are in the range of 50-90%. However, we should not preclude the possibility that more advanced methods of microwave generation could make very broadband emission efficient. In addition, larger bandwidth allows larger data transmission rates. Could ~100 MHz signals be broad transmission channels, a galactic information superhighway? Bandwidth alone is not necessarily a pulsar/beacon separator. (Note also that some pulsar observation is done with narrower bandwidths, so the true bandwidth is not observed, but is assumed to be broad, but may possibly not be.) SETI radiators may well avoid such a spectrum, to distinguish themselves from pulsars.

**Pulse length** Millisecond pulsars were observed later due to observational selection, e.g., integration time gives a natural bias against short periods. Cost-optimized beacons will likely be pulsed to lower cost, with a preference for shorter pulses due to source physics. On Earth, the higher source power, the shorter the pulse, due to breakdown physics, which is universal.

**Frequency** Pulsar searches cluster in the lower end of the microwave, but beacons may be more likely to appear at higher frequency. We expect cost-optimized beacons to appear at higher frequency (~10 GHz) due to the favorable scaling of cost with frequency.

**A Specific Case, PSR J1928+15**

As an example of analysis for SETI beacon possibilities, consider the transient bursting radio source, PSR J1928+15, which was observed in 2005 near the Galactic Disk at 1.44 GHz in a Arecibo 2 minute observation and not re-observed in 48 minutes of revisits[4]. (The candidate explanation in Deneva et al. is perhaps an asteroid falling into the neutron star from a circumpulsar disk, perturbing its magnetosphere.) Three pulses were received, the first and third down a factor of ten from the 0.180 Jy central pulse. Separation between was 0.402 sec (=1/2.48 Hz). The dispersion measure (DM) was 242 pc-cm$^3$, which places it at 24,000 ly, almost as far as galactic center (26,000 ly).

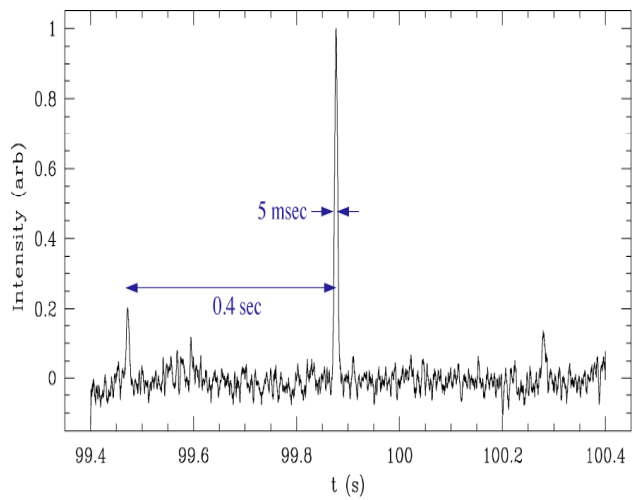

Dispersion of PSR J1928+15: De-dispersed time series showing three pulses, the center pulse with highest amplitude.

Its location is just below the galactic plane, as seen in Google Sky, where the pulsar is dead center in the microwave spectrum photo below. The plane can be seen to the right:

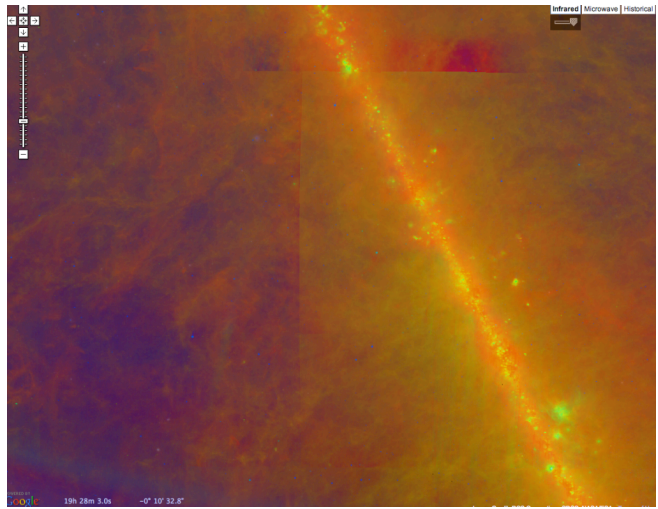

PSR J1928+15 is at the center of this photo in the microwave, just below the galactic plane

**Beacon Analysis**

Here we work exercises in understanding beacon tradeoffs, with a hypothetical PSR J1928+15 beacon. The methods discussed in reference 2 allow deduction of beacon parameters. Given a small set of observations, the principal parameters such as power and antenna area can be calculated. Then the time required to listen for a revisit of the scanning beacon can be made if we assume its search pattern.

To characterize a hypothetical PSR J1928+15 beacon, we make two working assumptions:

1) The beacon is a 'lighthouse' scanning the galactic plane. The source is a scanning beacon and, as it swept past, Arecibo caught the central pulse, the true beam. The first and third pulses are at the edges of the beacon beam width.

2) The beam bandwidth covers all channels of the 100 MHz span of the detector array. (The channel bandwidths are 0.39 MHz, with total BW 100 MHz.) This assumption drives the beacon power estimate. From the observed power density of 0.18 Jy, the total power across 100 MHz is $1.8 \times 10^{-19}$ W/m².

The scanning beam produces the three pulses 0.4 sec apart. In the next figure, we show that the pulses fit to a Gaussian beam pattern of width 0.5 sec, so the observed pulse heights are replicated.

Using the formulation of reference 2, and the earth cost parameters of 1 k$/m², 0.3 $/W, the cost-optimized parameters of a beacon producing $1.8 \times 10^{-19}$ W/m² at 24,000 ly are given in the Table.

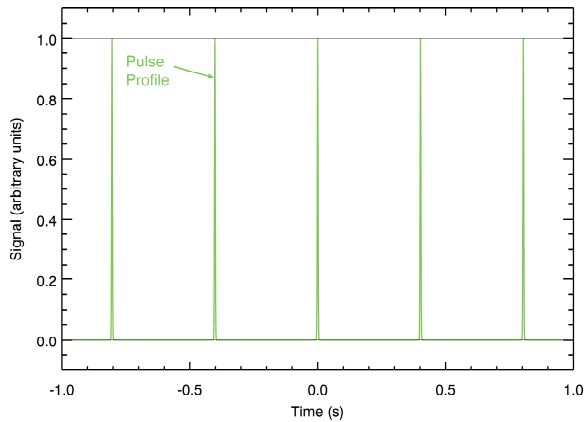
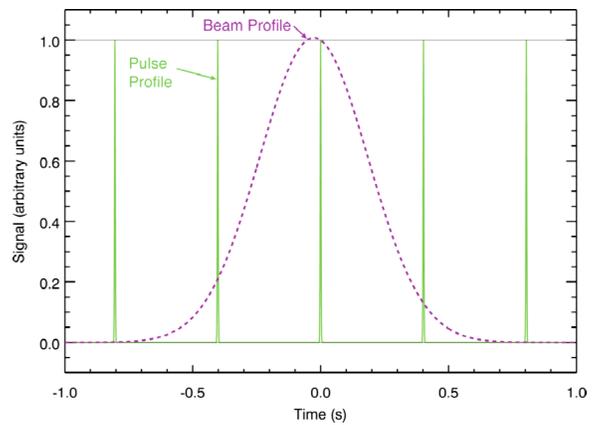
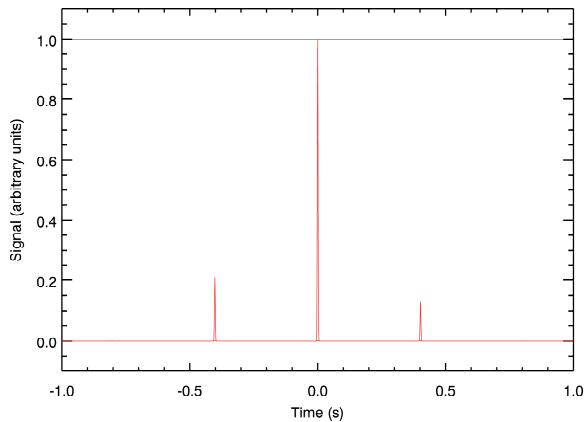
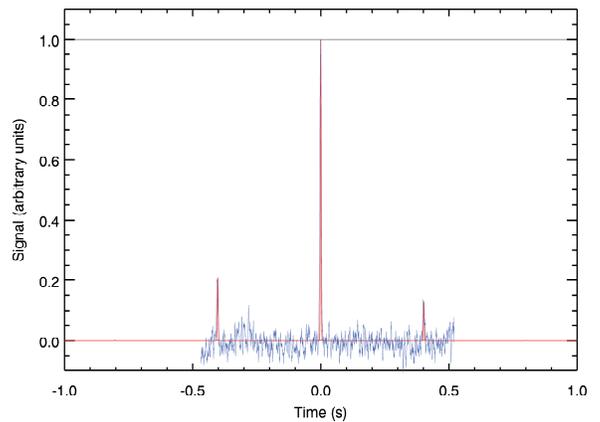

A hypothetical pulse profile, pulses the observed 0.4 seconds apart, convolved with a Gaussian beam shape 0.5 sec FWHM results in reduced first and third pulses and fits the observed signal.

# Beacon models producing the observed PSR J1928+15 signal

| Beacon | EIRP | Peak Power | Average Power | Antenna Diameter | Beam-width | Capital Cost | Operating Cost | Revisit time | K |
|---|---|---|---|---|---|---|---|---|---|
| | W | TW | GW | km | | B$ | B$/yr | | |
| A | $10^{23}$ | 1,600 | 3 | 25 | 9µrad | 980 | 3 | 6 years | 0.35 |
| B | $10^{21}$ | 190,000 | $4\ 10^{14}$ | 0.1 | 5mrad | $10^8$ | $3\ 10^{14}$ | 2 hrs | 0.86 |
| C | $10^{21}$ | 1,900 | 3,800 | 1 | 0.5mrad | 6,000 | 3,000 | 1 week | 0.66 |

*Beacon A: Cost-Optimized* Using cost-optimizing relations in reference 2 gives a beacon which has a big antenna, high peak power, but costs much less than the two other beacons discussed below. The small beam width gives a spot size of only 0.25 ly, meaning it's targeted to either ourselves or some target between (or behind) us. So, it is not a scanning beacon. If it were, the revisit time for a complete scan of the disk would be 6 years (see the calculation for the next example).

Total power in the spot area $A_s$ is then $P=S\ A_s=$ 1,600 TW, but the duty factor is low (5 ms/0.4 s=$2\ 10^{-3}$) so average power is 3 GW, about that of a standard nuclear installation. Note the *average* power, not the peak power matters. The beacon power system will store energy between bursts (or shots) in an intermediate store. Thus if it's a beacon, it comes from a civilization of our scale. On the Kardashev scale[5]

$$K \equiv \frac{\log_{10} P - 6}{10} = 0.35$$

where P is the beacon power. For comparison, Earth is K=0.73.

*Beacon B: Non-Cost-Optimized, small antenna* As an exercise in understanding beacon tradeoffs, assume the beacon antenna diameter is $D_t$=100 m. This leads to a small powerful beacon, with a large beam width

$$\theta = 2.44 \frac{\lambda}{D_t} = 5.1 \times 10^{-3} \text{rad}$$

where we use the observed frequency, 1.44 GHz. Then the spot size at our range is $R\theta$=122 ly.

The total power is then $P=SA_s$=190,000 TW, 10000 times the total electrical power of Earth. Thus if it's a beacon, it comes from a much more advanced and powerful civilization. On the Kardashev scale K=0.86, less than the power of an entire planet (K=1). Therefore, this beacon is an affordable luxury only to a civilization substantially in advance of us.

From the figure, the beam passing Arecibo in an interval 0.5 sec, so $d\phi/dt$ = 5.1 mrad /0.5 sec = $10^{-2}$ sec. If the pattern of the beacon is scanning the disk of thickness h, which is ~1300 ly at our location range of 24,000 ly, then the spot is moving at R $d\phi/dt$ = 250 ly/sec. The time for a cycle around the galactic circumference is $2\pi/[d\phi/dt]$ = 10 minutes. The number of such strips in the scan is 1350ly/122ly = 11. So the beacon will return in 11 x 10 minutes ~2 hours. It's understandable that 48 minutes of revisits hasn't seen it again in 40% of the revisit time. Of course, it could be scanning a smaller area, so that the revisit time would be sooner.

*Beacon C: Non-Cost-Optimized, large antenna* Assume $\underline{D_t=1 \text{ km.}}$ The beam width is reduced by a factor of 10 to 5 x $10^{-4}$ rad. Spot size diameter falls to 12ly. Power in the spot falls to 1900 TW, 100 times earth's entire power. The Kardashev scale falls to K=0.66. This is a civilization of planetary scale, commanding the entire energy of civilization of the previous example.

The spot moves at the same rate, 30 ly/sec. But since the spot is smaller, the number of strips in the scan increases to 1350ly/12.2ly = 110. So the beacon will return in 110 x 5 x $10^3$ sec = 5.5 x $10^5$ sec = 150 hours. Observers have revisited the site for 48 minutes, only 0.5% of the revisit time, and haven't seen it again.

**Discussion**
Note that the lower power beacons, i.e., lower on the Kardashev scale, will have a narrower beam, revisit less frequently, and so quite generally will be harder to observe.

The costs of beacons B and C are much larger than the cost-optimized beacon A, which is based on earth costs. For B and C, with their much smaller antennas, to be an economic choice, their civilizations must have very cheap power or much more expensive antenna costs relative to ours.

The beacon nomograph from reference 2 can be used to display the above examples. They all cover the same fraction of the sky, the galactic disk, but have very different dwell and revisit times.

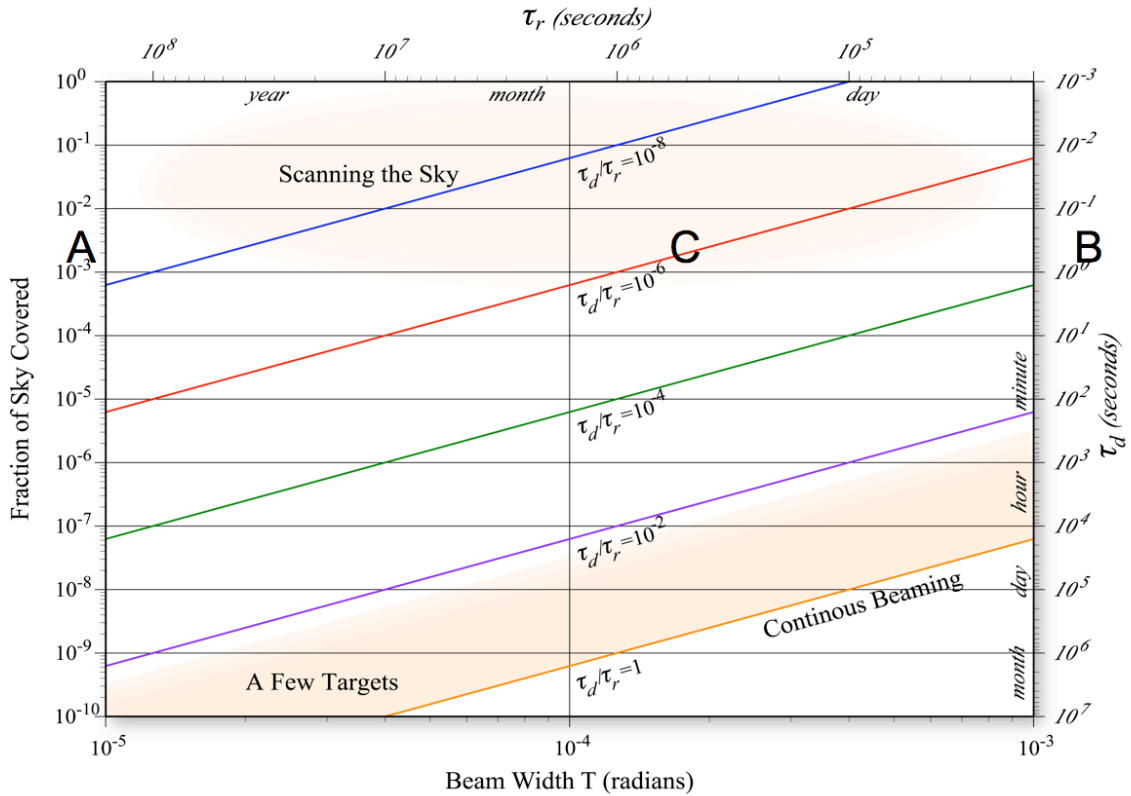

Figure 7. The three beacons of the Table displayed on the beacon broadcast strategy nomograph from Reference 2. A Beacon builder choosing values of θ and a sky fraction F to illuminate gives lines of constant duty cycle ratio (dwell time/revisit time, $\tau_d/\tau_r$) for the Beacon observer. Then right and top axes give ranges of these times for fixed $\tau_d/\tau_r$ ratio, and can not be correlated to the lower and left axes. The two relations are independent of each other, except in that they produce the same time ratios. Cost-optimal Beacons lie in the upper region, continuous Beacons targeting specific star targets are in lower region, can be observed with surveys observing for short times.

**Conclusions**

Galactic-scale beacons require resources larger than earth presently has available. From the examples, a civilization lower on the Kardashev scale will have a narrower beam, revisit less frequently, and so *will be harder to observe*. But lower power beacons will probably be more numerous. So we should learn how to identify them.

The discussion here shows a method of analyzing an observed radio transient in terms of a possible beacon. We urge observers to consider SETI beacons as a candidate explanation when perplexing non-repeating signals are seen in the radio sky.

Observers seeing brief pulses could do analysis like that done here, to estimate plausible revisit times. For example, always taking the cost optimized values and assuming a broadcast strategy (sky fraction in Figure 7) leads to a clear predicted revisit time. This can be useful in organizing future attempts to repeat transient observations.

We thank Gregory Benford for useful discussions.